# Molecular origin of blood-based infrared fingerprints

Liudmila Voronina[1,2,*], Cristina Leonardo[1,2], Johannes B. Mueller-Reif[3,ƚ], Philipp E. Geyer[3,4,ƚ], Marinus Huber[1,2], Michael Trubetskov[2], Kosmas V. Kepesidis[1], Jürgen Behr[5], Matthias Mann[3,4], Ferenc Krausz[1,2,6], Mihaela Žigman[1,2,6,*].

1. Department of Physics, Ludwig Maximilian University of Munich, Garching, 85748 Germany;

2. Max Planck Institute of Quantum Optics, Garching, 85748 Germany;

3. Department of Proteomics and Signal Transduction, Max Planck Institute of Biochemistry, Martinsried, 82152 Germany;

4. Novo Nordisk Foundation Center for Protein Research, Faculty of Health Sciences, University of Copenhagen, Copenhagen, 2200 Denmark;

5. Comprehensive Pneumology Center, Department of Internal Medicine V, Clinic of the Ludwig Maximilians University Munich (LMU), Member of the German Center for Lung Research;

6. Center for Molecular Fingerprinting, Budapest, 1093 Hungary.

ƚ New Address: OmicEra Diagnostics GmbH, Planegg, 82152 Germany

* Corresponding authors. Address: Am Coulombwall 1, 85748 Germany. Phone: +4917629522131

**Emails:** liudmila.voronina@mpq.mpg.de, mihaela.zigman@mpq.mpg.de

**Author contributions:** L.V. and M. Ž. designed the research plan; M. Ž., F.K. and M.M. initiated and led the study plan; J.B. led the clinical study; L.V., C.L., J.M., P.G. and M.H. performed the measurements; L.V., C.L., J.M., M.T., K.K. analyzed the data; L.V., M. Ž., J.M.-R. and F. K. wrote the paper.

**Competing Interests:** The authors declare no competing interests.

**Classification:** Physical/Biological sciences: Biophysics and computational biology.

## Significance

Previous studies demonstrated that infrared absorption spectra of blood sera may help disease detection. For clinical translation of this approach, it is important to determine the molecular origin of disease-related spectral perturbations. To that end, we supplemented infrared spectroscopy with biochemical fractionation and proteomic profiling, which provide detailed information about blood serum composition. We built a model to describe serum absorption based on the concentrations of the highly-abundant proteins and applied this framework to lung cancer detection. We find that it is the levels of acute-phase proteins that change most in the presence of the disease and generate its infrared signature. These findings inform future clinical trials and establish a framework that could be applied to probing of any disease.

## Abstract

Fourier-transform infrared (FTIR) spectroscopy of liquid biopsies combined with machine learning is a rapid and cost-effective approach carrying potential to aid biomedical diagnostics. However, the molecular nature of disease-related changes of infrared molecular fingerprints (IMFs) remains poorly understood, impeding the method's clinical applicability. Here we probe 148 human blood sera and reveal the origin of the observed variations in the infrared absorption spectra. We develop a workflow for reproducible crude chemical fractionation and evaluate the spectroscopic contribution of each fraction. Furthermore, we examine human sera with mass spectrometry-based proteomics and reconstruct the IMFs of blood sera as a linear combination of absorption spectra of the most abundant proteins. Using therapy naïve non-metastatic lung cancer as an example for a medical condition, we demonstrate that the disease-related differences in IMFs of blood sera are dominated by contributions from the protein fractions, rather than metabolites. More specifically, the major differences in IMFs between the lung cancer and the reference group are explained by the changes in concentrations of twelve identified proteins. Next to providing a new omics framework for systematically annotating IMFs with the so far missing molecular identity at gene/protein level, this study brings to light a combination of previously known highly abundant proteins – that, if used in a combined fashion, may be useful for detecting lung cancer. Altogether, tying proteomic to spectral information and machine learning offers clinically valuable understanding of disease-related changes in FTIR spectra of human blood serum.

## Introduction

Infrared spectroscopy is a well-established method of studying chemical substances *via* analyzing the vibrational transitions that are characteristic of their molecular structure (1). In particular, infrared molecular fingerprinting of human biofluids has the potential to provide information about the health state of individuals when combined with appropriate machine learning algorithms (2–13). The idea behind is to record an infrared absorption spectrum of the whole molecular ensemble composing blood serum using Fourier-transform infrared (FTIR) spectroscopy and pinpoint the deviations, associated with a given pathophysiological condition. However, the molecular origin of such changes in infrared molecular fingerprints (IMFs) is poorly understood (14, 15). The interpretation of the infrared absorption spectra is currently largely restricted to the characteristic spectral signatures of various functional groups (16–18). However, these are contained in many different types of biomolecules, their spectral features in aqueous environment are broad and strongly overlapping, and the molecular complexity of biofluids is extremely high. Therefore, the understanding of the underlying molecular changes of the IMFs has so far been limited (19, 20).

Thorough exploration of the molecular origin of IMFs would be instrumental for successful application and verification of molecular fingerprinting in clinical settings (7). It would allow for improved sample preparation, ensure that the spectral features used for building the computational models are indeed caused by a medical condition and not by confounding factors and help define the possible limitations of blood-based IMFs' applicability (21). To that end, several studies measured the concentrations of a range of analytes in human blood serum using conventional biochemical methods and demonstrated that IMFs can be used to retrieve these concentrations using multivariate regression or consecutive spectral subtraction approaches (8, 22–26). However, they come up short in determining how exhaustive the list of molecular constituents is and connecting disease-related changes in the molecular composition of biofluids to the changes in the corresponding IMFs (25).

It had been suggested that large variations in blood-based IR spectra may be caused by a varying albumin-to-globulin ratio (27). Indeed, the spectroscopic signature of human blood serum is vastly dominated by a few highly abundant molecular components, such as human serum albumin (HSA) and immunoglobulins (28). To overcome the challenge of strong molecular signals that overshadow the signals from less abundant molecules, splitting complex biological samples into several fractions of different chemical nature is beneficial (29, 30). Previously, ultrafiltration has been used to fractionate human blood serum based on molecular weight of the components (14, 23, 31, 32). However, these methods introduce unwanted chemicals in non-reproducible fashion (33). In this study, we chose to adapt a combination of solvent-extraction sample preparation protocols, which are typically used in metabolomics (34) and proteomics (35), because of their robustness and speed (36).

In order to explore the dependence of the IMFs of human blood serum on its molecular composition, spectroscopic molecular fingerprinting should be ultimately combined with a technique that is able to provide molecular-specific information over a high dynamic range (37). Recently, a high-throughput mass spectrometry (MS)-based proteomic workflow has been established for the analysis of human blood plasma profiles (38). We adapted this technology for human blood serum and applied it to our sample set in order to model the IMFs of hydrated biofluids as a linear combination of molecular components. Such a parallelized FTIR-MS approach for molecular annotation of disease-relevant vibrational fingerprints of human blood derivatives has been lacking this far.

With the gained understanding of the molecular composition underlying the IMFs of human blood serum, we compare the samples of lung cancer patients (TNM clinical stages II and III) with reference individuals matched in age, gender and smoking status. The ability of FTIR spectroscopy of blood serum to discriminate lung cancer cases from controls has been previously shown in several studies (39, 40). Pattern recognition algorithms were used to identify non-small cell lung carcinoma and subtype the disease conditions (39). Independently, the ratio between intensities at 1080 and 1170 $cm^{-1}$ was put forward as the most informative for disease detection, and it was suggested that changes in the protein secondary structure might be correlated with lung cancer (40). Other types of cancer have also been detected with various efficiencies using blood-based IMFs, with little insight into molecular changes for the reasons stated above (13, 41–45).

In this study, we obtain reproducible, cost- and time-efficient IMFs of human sera and use proteomic measurements to facilitate their understanding at a molecular-level. In particular, we reveal a pattern of changes of human blood serum composition, which correlates with the presence of lung cancer and results in an observable difference between IMFs of blood sera of lung cancer patients compared to the reference group. Both spectral and molecular information was used to build explainable classification models for lung cancer detection (46). This paradigm can be applied to other health phenotypes in order to develop reliable and transparent diagnostic tools.

# Results

## 1. Decomposing complexity of human blood sera using biochemical fractionation

We recorded infrared absorption spectra of liquid human blood sera in the range from 1000 to 3000 $cm^{-1}$. The spectra are dominated by amide bands that are attributed to the vibrations of protein backbone (47). In particular, the most prominent feature between 1600 and 1700 $cm^{-1}$ (Amide I band) is characteristic of the secondary structure of the proteins (47). The region on the red side of the spectrum (1000-1200 $cm^{-1}$) is often referred to as “carbohydrate region”, because of the typical absorption patterns that glycans exhibit here (17). Finally, lipids produce several absorption bands around 1735 $cm^{-1}$, 2852 $cm^{-1}$ and 2926 $cm^{-1}$ (48).

Attributing the distinct features of the mid-infrared absorption spectrum of human blood serum to a specific molecular class is somewhat oversimplified, since absorption spectra of various biological molecules often overlap. In order to gain deeper insight into the origins of different spectral features, we built a comprehensive model of the human blood serum absorption. To this end, we used a set of 148 prospectively collected blood serum samples (Fig. 1*A*).

As a first step, we recorded the IMFs of each full intact, fluid, serum sample using high-throughput automated FTIR spectrometer in transmission mode (black line in Fig. 1*B*, Dataset S1) (4). Next, we biochemically fractionated each sample into three fractions and recorded their IMFs (colored lines in Fig. 1*B*) in order to assess the relative contributions of roughly defined molecular classes, i.e. proteins and metabolites. In parallel, we used proteomic analysis of the crude sera and human serum albumin (HSA)-depleted fractions to characterize the efficiency of HSA depletion and the molecular composition of each protein fraction.

**Fig. 1.** Decomposing complexity of human blood sera using chemical fractionation. (A) Overview of the workflow of the study. (B) Average infrared molecular fingerprint (IMF) of human blood serum of 93 reference individuals and the corresponding IMFs of 3 fractions. The dashed vertical line shows the position of the Amide I band in the HSA-enriched fraction. The two lower inserts highlight the regions with the largest relative differences between the fractions. (C) Reproducibility of the fractionation protocol assessed with proteomic and FTIR measurements. Left axis: coefficients of variation for the levels of 12 proteins considered in this study for the same 8 serum samples with and without fractionation as well as their between-person variability in 93 control individuals. Right axis: the corresponding variations in the IMFs, averaged across wavenumbers.

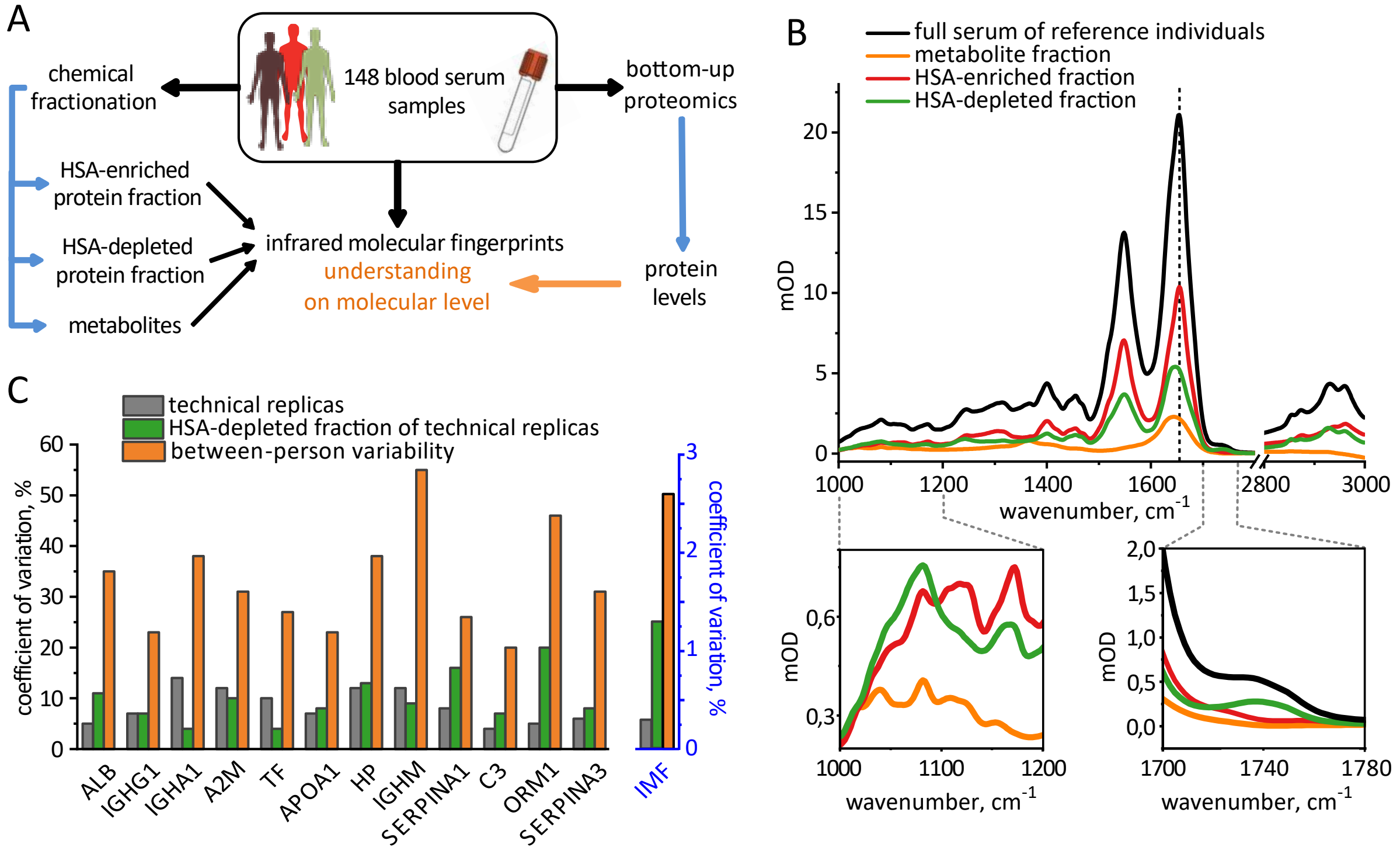

Human serum albumin is the most abundant serum protein and constitutes about a half of total protein mass (28). It is helpful to separate HSA away from other proteins, because its intense absorption potentially obscures the signals from other molecules (29). For this purpose, we first precipitated most of the proteins using cold ethanol (35). The supernatant was enriched in HSA, which we precipitated in the next step to separate it from metabolites (49). All three fractions (HSA-depleted proteins, HSA-enriched proteins and metabolites) were dried in vacuum and re-dissolved in water prior to the spectroscopic measurements.

We assessed the reproducibility of our fractionation protocol both with FTIR spectroscopy and proteomic analyses (Fig. 1*C*). First, we estimated the measurement uncertainty of the proteomic workflow as the coefficient of variation (CV) in repeated measurements of the same single human blood plasma sample. The average CV for the 12 proteins considered in this study (see below) in the crude plasma samples is 9 %, and it rises to 10% in the HSA-depleted fraction of the same sample, suggesting that the process of fractionation adds only minor error compared to the instrumental one. The CV measured for 93 reference individuals provides a rough estimate for the between-person variability, which is higher than the instrumental error for all considered proteins (33 % on average). The analysis based on IMFs leads to similar conclusions (Fig. 1*C*, right axis).

We further compared the spectral intensities of each of the fractions (Fig. 1*B,* Dataset S2). This procedure facilitates several unexpected conclusions about the nature of the IMFs of crude blood sera: Firstly, the signals between 1000 and 1200 $cm^{-1}$ are typically attributed to carbohydrates (16). Indeed, we detected the metabolite fraction containing free carbohydrates, exhibiting characteristic pattern in this region of the spectra. However, the intensity of the signals from both two protein fractions combined is an order of magnitude higher than that of metabolite fraction in this spectral region. We attribute this effect to glycosylation of proteins and further demonstrate it below. Additionally, we show that around 10% of the intensity of the Amide I band (1654 $cm^{-1}$ in crude serum), which is typically attributed to proteins (16), is actually contributed by metabolites.

Altogether, our fractionation workflow enabled us to disentangle the quantitative contributions of metabolites and proteins to the IMF of crude blood sera. Since the absorption of proteins fractions is, as expected, significantly higher than that of metabolites, in the next step we focused on understanding and modeling the contribution of protein absorption to the overall fingerprints.

## 2. Towards molecular understanding of infrared fingerprints using proteomics

We demonstrated that the IR spectrum of blood serum mostly exhibits signals originating from the protein absorption. It is therefore important to understand how various proteins of blood sera contribute to the overall IR absorption spectra of this biofluid. To that end, we performed bottom-up proteomic analysis of the same samples (Dataset S3). They were subjected to an established mass-spectrometry based proteomics pipeline (38). In brief, proteins in the sample are denatured and disulfide bonds reduced and quenched. Proteins are then digested into tryptic peptides and desalted. The peptides are separated by reversed phase chromatography coupled online to the mass spectrometer to detect the mass to charge ratios of peptides and their fragments in a quantitative manner. This enables software-dependent peptide identification and subsequently quantitative protein assembly from detected peptides (50, 51).

The first ten proteins listed in Fig. 1*C* are the ten most abundant proteins in human blood serum (SI Appendix Table S1). The quantitative values for each protein (so called 'label-free quantification' or LFQ values) provided by proteomic measurements are suited to characterize the differences between subjects in a study, but not directly proportional to the absolute concentrations of proteins (52), as revealed by Table S1 (SI Appendix). To obtain the actual protein concentrations, we re-scaled the LFQ values using the average reference concentrations of these proteins in healthy subjects.

To be able to link the actual individual protein levels directly to the IMFs of blood sera, we measured IR absorption spectra of each of the 10 most abundant proteins separately, dissolved in phosphate-buffered saline (PBS). Fig. 2*A* demonstrates the IR spectra of 5 highly abundant proteins (SI Appendix Fig. S2 for all proteins). The position and shape of the Amide I band is characteristic for their secondary structure and qualitatively corresponds to the known β-sheet and α-helix content of proteins (47). As expected, alpha-1-acid glycoprotein (ORM1 in Fig. 2*A*) shows particularly high absorption in the region of 1000-1200 $cm^{-1}$, because about 45 % of its dry mass is comprised of carbohydrates (53).

**Fig. 2.** Molecular modeling of infrared fingerprints based on serum proteomic profiling. (A) Examples of infrared absorption spectra of human serum proteins at the same concentration, 5 mg/mL. (B) Average IMF of 148 human blood sera, each modelled as a sum of contributions of 10 proteins compared to the average experimentally measured IMF. (C) Average vector distance between the model and experimental spectra for all 148 samples depending on the number of proteins introduced into the model.

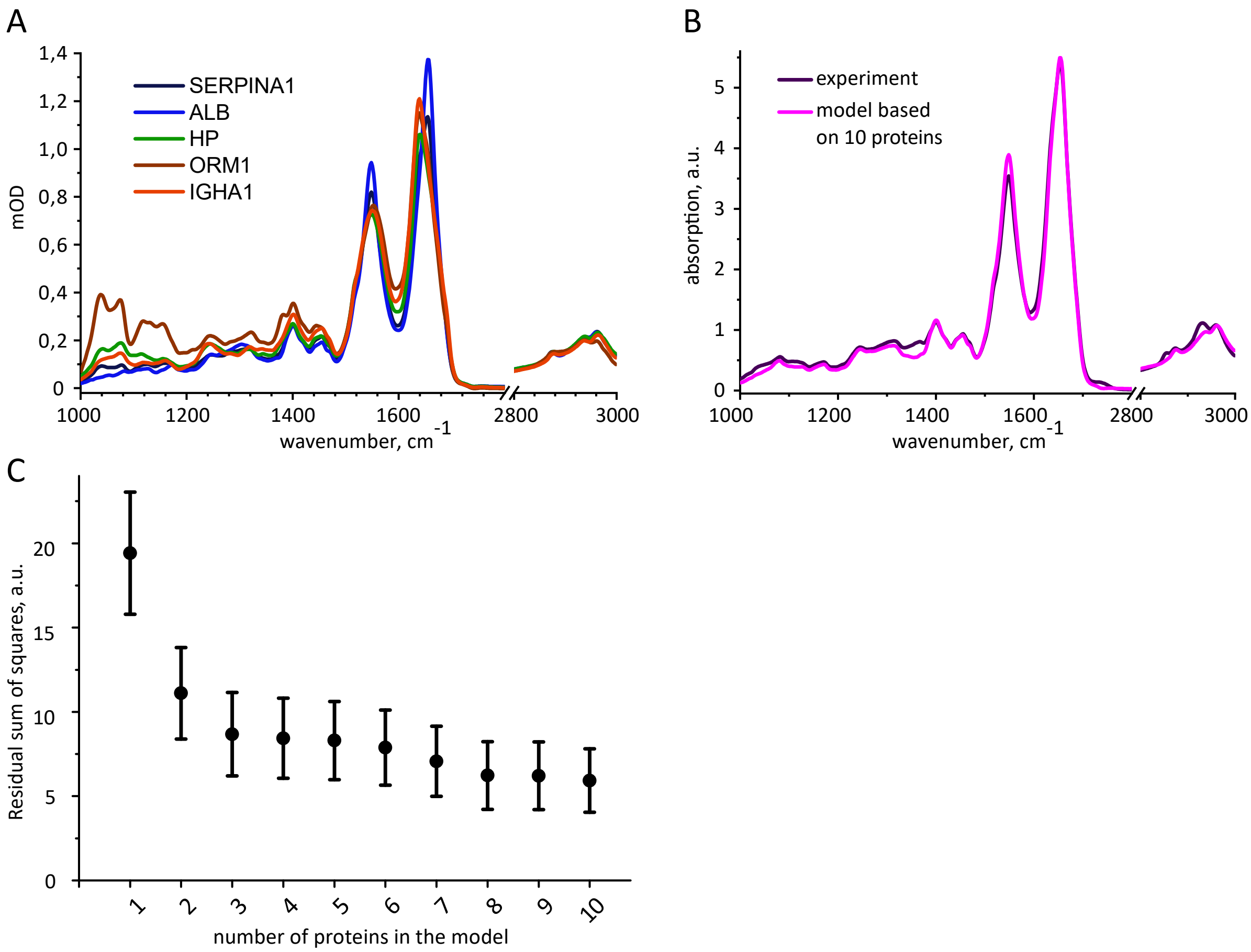

In order to estimate the contribution of each protein to the IMF of blood serum, we modeled the absorption spectra of every individual's serum as a sum of IR absorption spectra of proteins multiplied by their respective concentrations, measured by proteomics:

$$IMF(\tilde{\nu}) = \sum_i C_i * S_i(\tilde{\nu}) ,$$

where $\tilde{\nu}$ represents wavenumber, $C_i$ – concentration of the protein *i* in mg/mL, $S_i(\tilde{\nu})$ – absorption spectrum of the protein *i* for 1 mg/mL.

We started by taking into account the spectral contribution of HSA only (*i*=1) and building complexity by adding proteins one by one, in the order as listed in SI Appendix Table S1. Fig. 2*C* shows how the model becomes closer to the experimentally measured IMFs with every additional protein. Adding further lower abundant proteins to the model is expected to yield only small improvements, since the total concentration of remaining proteins that are beyond the ten molecules considered here is about the same order of magnitude as the level of complement component C3.

In Fig. 2*B* we compare the average modeled and experimental absorption spectra of human blood serum. Given the linear character of the model and the limited number of considered components, the matching is remarkably high. The only prominent peaks missing from the modeled spectra are the C=O (at 1735 $cm^{-1}$) and C-H stretches (at 2852 $cm^{-1}$ and 2926 $cm^{-1}$) known to be unique for lipids (48). Indeed, the average concentration of cholesterol in human blood serum is of the same order of magnitude as the last proteins we considered (54). The model can, therefore, be further refined by including cholesterol and other metabolites, such as ATP, melanin, glucose and urea. In fact, adding the entire metabolite fraction to the model further reduces the RSS between the model and the experiment by 50 % (SI Appendix Fig. S3).

### 3. Combining MS-based proteomics and IR fingerprinting reveals lung cancer-related molecular changes in blood serum

Having obtained a simple model of the IR absorption of human blood serum, we can address the question how this absorption changes as a consequence of a disease. In this study we focused on lung cancer, as the most common cause of cancer-related deaths worldwide (55). We compare the IMFs of prospectively collected sera between two cohorts: 55 lung cancer patients (therapy naïve, prior to any cancer-related therapy, at TNM clinical stages 2 and 3) with 93 reference individuals. In the latter cohort we gathered non-symptomatic individuals (“healthy”), patients with chronic pulmonary obstructive disease (COPD) and individuals with lung hamartoma, to challenge our detection regime by non-cancerous lung diseases. Importantly, the cohorts are gender, age and smoking-status matched (SI Appendix Table S2).

We find that infrared molecular fingerprints of lung cancer patients clearly differ from that of reference individuals. The black line in Fig. 3*A* shows the difference between the average IMF of lung cancer patients and those of references as a function of wavenumber, which we specify as “differential fingerprint”. The p-values of the most prominent spectral peaks are below $10^{-6}$ (SI Appendix Table S3), strongly suggesting that the differences between the IMFs of two cohorts are statistically significant. To further quantify these differences, we applied support vector machine (SVM) algorithm to classify the samples into two classes – cancer cases and reference individuals. To that end, the data were split into train and test sets, employing 10-times repeated 10-fold cross-validation. The area under the curve (AUC) of the receiver operating characteristics (ROC) curve was used as a measure of classification efficiency. For the classification of lung cancer patients versus references, the model reveals an AUC of 0.85±0.1, implying that the SVM model can, in principle, be trained to distinguish between the two cohorts.

We find that the differential fingerprint of lung cancer has a specific shape, with prominent features around 1000-1200 $cm^{-1}$, as well as in the Amide I and Amide II regions. Such shape could result from alternations in the proteins secondary structure, as previously suggested (40) or, alternatively, from the changes in their concentration (21). The distinction between the two possibilities can only be obtained by comparison of two sample sets with a technique that provides information about molecular concentrations.

The HSA-enriched and HSA-depleted fractions reflect the largest differences between lung cancer and reference samples with p-values below $10^{-6}$ (SI Appendix Table S3), while the metabolite fraction is not significantly different in the samples from reference individuals versus these of the lung cancer patients. This finding is confirmed by the AUC values: for the metabolite fraction the AUC is 0.62±0.2, while for the HSA-enriched fraction it is 0.82±0.1, and for the HSA-depleted fraction - 0.75±0.1. Thus, we turned to the proteomic measurements of the same sample set - aiming for the identification of individual proteins responsible for the observed changes in the IMFs.

**Fig. 3.** Lung cancer-related molecular changes in blood serum, based on comparison between 55 lung cancer patients and 93 reference individuals. (A) Differential fingerprints of lung cancer in full sera: experimentally measured and modeled based on the levels of 12 proteins. The shaded area shows the standard deviation of the IMFs of the reference group. (B) Change in the concentrations of proteins in blood serum caused by lung cancer, measured by proteomics. The proteins are ordered according the absolute difference in the concentrations in lung cancer and control individuals. *- p-value below 0.05, ** - p-value below 0.0005, ***- p-value below $10^{-6}$, no star – p-value above 0.05. (C) ROC curves based on the experimental measurement of IMF of full serum and the set of 12 proteins measured by proteomics. The STDs are 0.1 for AUC in panels (C) and (F). (D) Differential fingerprints of lung cancer in HSA-enriched fraction: experimentally measured and modeled based on the levels of 3 proteins. (E) Change in the concentrations of proteins in HSA-enriched fraction caused by lung cancer, measured by proteomics. (F) Comparison between the ROC curves based on the experimental measurement of IMF of HSA-enriched fraction and the corresponding set of 3 proteins.

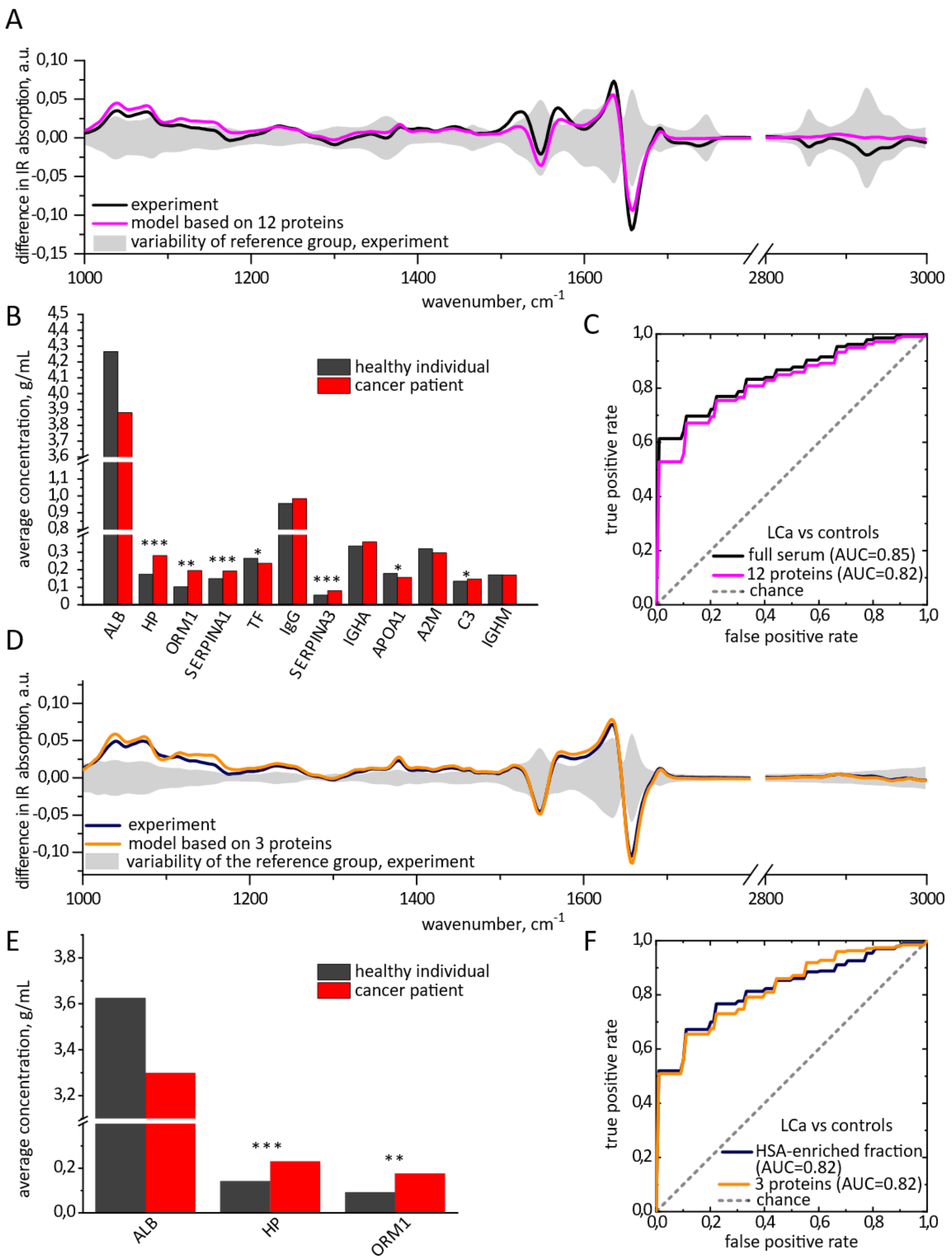

In line with previous research (56–60), we find a number of proteins that demonstrate p-values below 0.0005 (SI Appendix Table S4). However, the purpose of this study is not the search for specific biomarking candidates; instead, we wish to evaluate whether lung cancer results in a pattern of changes in protein concentrations responsible for its IR signature.

The first question we have addressed is: which proteins do we have to consider in order to model the differences in the IMFs between the lung cancer patients and reference individuals. The *differential* fingerprint is affected by the disease-related *absolute change* in the protein concentration due to the linear character of the absorption measurement. Therefore, we ranked all detected proteins according to the absolute difference in average concentration between lung cancer and reference samples, as measured by MS (SI Appendix Table S5).

Out of ten proteins that are most extensively changing, eight are also among the ten most abundant proteins in the blood sera. We further identify other proteins reflecting the differences between the two sample sets, such as alpha-1-acid glycoprotein-1 and alpha-1-antichymotrypsin: although their concentrations in non-symptomatic subjects are below the ten most abundant proteins, they are changing significantly in lung cancer patients and thus have to be taken into account to accurately model the disease differential fingerprint. In total, we considered twelve proteins for the model of lung cancer differential fingerprint, as shown in Fig. 3*B*: ten most abundant ones and two additional ones that are changing most significantly.

After we have modelled the IMF of every individual as described above, the differential fingerprint of lung cancer was calculated as the difference between the average fingerprint of lung cancer patients and reference individuals. The resulting curve of this twelve-protein model very closely resembles the measured differential fingerprint, reflecting all the important features (pink line in Fig. 3*A*). Moreover, the binary classification of lung cancer cases versus reference individuals based on the concentrations of the twelve identified proteins produces an AUC of 0.82±0.1, which is close to the value for experimentally measured serum spectra (0.85±0.1). These findings suggest that most of the information in IMFs regarding lung cancer status stems from the molecular changes in these twelve proteins. Moreover, such kind of information can be measured in time- and cost-efficient manner by applying FTIR, without the need to measure the concentrations of each of the protein separately.

Interestingly, the three proteins that change the most between the lung cancer patients and the reference group (namely, HSA, haptoglobin and alpha-1-acid glycoprotein 1, Fig. 3*B* and 3*E*) remain predominantly contained in the HSA-enriched fraction during the fractionation procedure.  This explains the high AUC obtained for this protein fraction: 0.82±0.1, blue line in Fig. 3*F*. It further suggests that most of the molecular information about the presence of lung cancer is encoded in the concentrations of the three proteins named above, out of all twelve proteins analyzed. Indeed, the SVM binary classification based on the concentrations of these three proteins reveales the AUC of 0.82±0.1, the same as based on all 12 proteins considered above.

We modeled the IMFs of the HSA-enriched fraction as detailed above, taking into account the proportion of each protein in HSA-enriched fraction compared to full serum (SI Appendix Table S1 and Fig. S1). In line with only a minor contribution of low-abundant proteins and metabolites to the IR spectra of HSA-enriched fraction, we find that the model very well reproduces the experimental curve (Fig. 3*D*).

In summary, we observe statistically significant differences between the IMFs of blood serum of lung cancer pateints compared to reference individuals. Biochemical fractionation and proteomic profiling of the very same sample set helped identify the compounds responsible for these differences and revealed previously unappreciated pattern of changes in the concentrations of well known proteins that we find to be characteristic of lung cancer.

## Discussion

Although FTIR has been used over decades and blood-based studies suggested the applicability of this approach to disease diagnostics, the molecular nature of blood-based infrared molecular fingerprints (IMFs) and changes therein has not been well understood. Being cost- and time-efficient, suitable for high-throughput approaches, IMFs could greatly contribute to clinical practice if their robust correlation with any given condition is reproducibly demonstrated. Molecular understanding of the IMFs along with computational models may open up a path towards informed choice of biofluid (e.g. serum *vs* plasma), improved sample preparation and possibly even initial steps of the biomarker identification. Here we took advantage of a prospective clinical study and examined the samples with two independent techniques - IR spectroscopy and mass spectrometry (MS)-based proteomics - with the goal to elucidate the molecular entities dominating human blood-based IMFs.

As a first step to decompose chemical complexity of IMFs, we established a protocol for highly-reproducible fractionation of crude human blood sera into three fractions: human serum albumin (HSA)-enriched proteins, HSA-depleted proteins, and metabolites. The strongest IR absorption signal in human blood serum arises from proteins. We therefore measured their relative concentrations in the samples using MS-based proteomic profiling and used the concentrations of ten most abundant proteins to reconstruct individual spectra of the human blood serum. The model can be further developed by adding highly abundant metabolites and additional proteins until the model reproduces measured IMFs within their noise limit. In particular, it has been shown previously that in addition to the proteins discussed here, FTIR spectra of blood plasma provide information about the levels of lactate, urea, apolipoproteins B and C, as well as immunoglobulin D (25). However, the data presented here suggest that our 10-protein-based approach leaves little room for improvement in modelling IMFs measured by FTIR spectroscopy. The ultimate limitation of such modeling lies in the linearity of the model, disregarding any interaction between different blood components.

Infrared molecular fingerprints acquired by field-resolved spectroscopy (61) may drastically increase the precision of infrared molecular fingerprinting by reducing the noise limit. This will render smaller molecular contributions significant, uncovering thereby more molecular information just as the combination of further biochemical fractionation (e.g. by liquid chromatography) with field-resolved spectroscopy will do. Both may allow more lower-abundance molecules to contribute to the identification of a pathophysiological condition.

In this study we use lung cancer as a case scenario of a medical condition, the outcome of which could significantly benefit from early detection. We find that IMFs of sera samples of lung cancer patients differ significantly from that of reference individuals. Using MS-based proteomics, we identify a pattern of known highly-abundant proteins that determine the observed change in the IMFs of blood sera. Some of them have been previously linked to cancer: unexplained hypoalbuminaenia has been assosiated with increased cancer risk (62), and low pre-treatment albumin level – with poor survival rate (63). Moreover, in line with our findings, the levels of haptoglobin, complement component C3, alpha-1 antytrypsin and alpha-1-acid glycoprotein were previously shown to rise in blood of lung cancer patients (57–59).

Importantly, although these proteins are not specifically challenging to detect and measure, they have previously not been used in a combined fashion to help detect or diagnose lung cancer. It is meanwhile widely accepted that using multiple biomarking molecules together, as a pattern, is more effective and robust for detecting a particular health condition (28). Infrared fingerprinting of human blood serum takes this approach to a new level: here we effectively combine a wide range of molecules into a single IR spectrum, that can be easily measured and interpreted. To illustrate that, we considered the levels of all 114 proteins detected by proteomics in every sample. Importantly, the binary classification efficiency based on all these proteins measured separately is not higher than the efficiency based on a single IMF measurement (SI Appendix Table S6).

Lung cancer induces a number of changes in the levels of blood serum proteins that have been previously linked to acute-phase response, and it is well-known that cancer is often associated with inflammatory states (64, 65). In line with the general discussion in the field (21), our findings underscore the need for additional clinical studies that would look into the specificity of IMFs. A well-designed reference cohort should include individuals with potentially similar pattern of changes in the blood composition: for example, in the case of lung cancer, with chronic or acute inflammation, which is often accompanying malignant states. Due to cost-efficiency and rapidity of blood-based infrared molecular fingerprinting, it could still find a wide range of applications, even if its specificity proves insufficient for screening applications. Thus, general molecular-level understanding of the disease-related changes in IMFs will help establish better clinical study design, and ultimately lead to improved approaches to medical diagnostics.

The paradigm presented here could in principle be used for any pathophysiological condition. After having recorded the IMFs of patients and compared them to matched reference individuals, one could use biochemical fractionation to determine which molecular class is responsible for the disease-related differences and perform in-depth omics profiling of the identified fraction. This would provide insights into the nature of information that infrared molecular fingerprinting is able to provide and into its additional value compared to well-established clinical tests. Moreover, combining biochemical fractionation with field-resolved spectroscopy-based infrared molecular fingerprinting (61) might yield deeper molecular insight along with higher specificity and sensitivity for disease detection. Ultimately, the larger clinical studies with purposefully chosen reference groups, stratified and controlled for comorbidities, may bring IMF – a cheap and time-efficient method – closer to everyday clinical use.

## Materials and methods

### 1. Chemicals and reagents.

Methanol and ethanol of HPLC-grade, sodium chloride and proteins at highest available purity rate were purchased from Sigma Aldrich GmbH (Taufkirchen, Germany). The proteins that were purchased as powder were diluted in 20mM PBS buffer (Sigma Aldrich GmbH). If traces of additional salts were present in the protein solution, the buffer was exchanged to 20 mM PBS using Nanosep Omega centrifugal filters with 3 kD cutoff (VWR, Germany).

### 2. Clinical study participants.

We performed a prospective clinical study on lung cancer, including subjects with benign conditions and non-symptomatic healthy volunteers as reference (see SI Appendix Section S1 for a list of involved clinical centers). All participants signed written informed consent form for the study under research Study Protocol # 17-182, approved by the Ethics Committee of the Ludwig-Maximillian-University (LMU) of Munich and performed in compliance with all relevant ethical regulations. Analyses focus on subjects with clinically confirmed carcinoma of lung at the TNM clinical stages II and III, with no metastases, prior to any cancer-related therapy, and without any other cancer occurrence. Healthy references were non-symptomatic individuals, without any history of cancer, not suffering from any cancer-related disease nor being under any medical treatment. Lung cancer cases were compared to matched individuals from the following groups: Chronic obstructive pulmonary disease (COPD), pulmonary hamartoma and non-symptomatic healthy individuals matched for gender, age and smoking status. Full breakdown of all participants is listed in SI Appendix Table S2.

### 3. Blood sample collection and preparation

Blood samples were collected, processed and stored using previously defined standard operating procedures: Blood draws were all performed using Safety-Multifly needles of 21G (Sarstedt AG & Co KG, Germany) into 4.9 ml serum tubes, centrifuged at 2.000 g for 10 minutes at 20 °C, aliquoted and frozen at -80°C within 3 hours from the time of sampling. All samples were thawed, further aliquoted for measurement and re-frozen at -80°C to ensure a constant number of freeze-thaw cycles before analysis. Before any measurement, the aliquots were thawed at room temperature, shaken for 20 seconds, and spun down again.

### 4. Fourier-transform infrared spectroscopy measurements

Measurements of liquid biofluids, their fractions and single proteins were all performed in hydrated, fluid state using an automated FTIR device MIRA-Analyzer (Micro-biolytics GmbH, Germany) with a flow-through transmission cuvette ($CaF_2$ with 8 µm path length), as demonstrated previously (4). See SI Appendix Section S2 for a detailed description.

The absorption spectra of single proteins were measured at 1 to 5 mg/mL concentration depending on the sample availability. The spectrum of PBS buffer was subsequently measured and subtracted from the protein spectrum prior to further pre-processing. When the buffer was exchanged using centrifugal filters, which implies sample loss, the resulting protein concentration was determined using BCA Protein Assay.

### 5. UPLC-MS proteomics measurements

Sample preparation was carried out according to our Plasma Proteome Profiling pipeline (38), employing an automated setup on an Agilent Bravo Liquid Handling Platform. Samples were measured using LC-MS instrumentation consisting of an Evosep One (Evosep, Odensee, Denmark), which was coupled to a Q Exactive HF-X Orbitrap (Thermo Fisher Scientific) using a nano-electrospray ion source (Thermo Fisher Scientific). Raw files were analyzed by MaxQuant software, version 1.6.3.3 (66), and peptide lists were searched against the human UniProt FASTA database. See SI Appendix Section S3 for further details.

**6. Fractionation of blood serum**

All the samples and all reagents were kept at $4^0$C during the process. The samples were processed in batches of 24, including 4 quality control samples per each batch (see above). Two-step fractionation of liquid biofluids has been performed. The goal of the first step is to separate most of the proteins from the suman serum albumin (HSA) and follows the previously proposed (35). To that end, 0.1M NaCl and 42% of ethanol have been added to the samples. They were vortexed for 1 hour, then centrifuged for 20 minutes at 16000 rcf, so that a pellet containing most of the proteins (HSA-depleted fraction) is formed. The supernatant that contains HSA and metabolites was transferred to a new tube, while the pellet was re-dissolved in water *via* vortexing for 1.5 hours. A small pellet was left in the tube. We have shown that if the HSA-depleted protein pellet is vortexed for longer, the left-over pellet is reduced, but the FTIR spectra of the supernatant are identical to those obtained after 1.5 hours. The HSA-depleted fraction has been transferred to a new tube and placed into the vacuum concentrator (Concentrator plus, Eppendorf GmbH, Germany) for 3 hours. To avoid clogging of the automated measurement system, the HSA-depleted fraction has been centrifuged for 15 minutes at 15000 rcf and frozen at $-80^0$C until further use.

To separate HSA-enriched protein fraction from metabolites, we added 59% of pre-cooled methanol, vortexed the samples for 1 minute and centrifuged for 15 minutes at 15000 rcf, so that a pellet containing HSA and other proteins was formed (49). The supernatant was transferred to a new tube and fully dried in the concentrator in 3 hours. The metabolites were then re-dissolved in water *via* vortexing for 2 minutes and frozen at $-80^0$C until further use.

The HSA-enriched pellet was fully re-dissolved in water *via* vortexing for 2 minutes and placed into the vacuum concentrator for 3 hours. The resulting HSA-enriched protein fraction was frozen at $-80^0$C until further use. The total time required to process a sample batch was 8 hours.

### 7. Classification models

The data analysis was performed using the *Scikit-Learn* (67) (v. 0.23.2) module in Python (v.3.7.6). We trained classification models based on linear support vector machines (SVM) algorithm – as implemented in the *LinearSVC* class with default parameters. We evaluated the model using stratified 10-fold cross validation, repeated 10-times with different randomization in each repetition. For the visualization of the model performance, we use the notion of the receiver operating characteristic (ROC) curve. As an overall metric for model performance, we use the area under the ROC curve (AUC). For the evaluation of the mass-spectrometry data, we performed ranking of individual proteins using forward feature selection based on the SVM-classification performance.

## Acknowledgements

We would like to thank Frank Fleischmann, Catherine Vasilopoulou, Jacqueline Hermann, Katja Leitner, Sigrid Auweter, Daniel Meyer, Beate Rank and Incinur Zellhuber for their help with this study. In particular, we wish to acknowledge the efforts of many individuals who participated as volunteers in the clinical study reported here. We also thank A. Barth for his insightful suggestions.